# Stable gigahertz- and mmWave-repetition-rate soliton microcombs on X-cut lithium niobate

Yunxiang Song[1,2,†], Xinrui Zhu[1], Xiangying Zuo[1], Guanhao Huang[1], Marko Lončar[1,†]

[1] John A. Paulson School of Engineering and Applied Sciences, Harvard University, Cambridge, MA 02138, USA
[2] Quantum Science and Engineering, Harvard University, Cambridge, MA 02138, USA
[†] *ysong1@g.harvard.edu*, *loncar@g.harvard.edu*

**Soliton microcombs are a cornerstone of integrated frequency comb technologies, with applications spanning photonic computing, ranging, microwave synthesis, optical communications, and quantum light generation. In nearly all such applications, electro-optic (EO) components play a critical role in generating, monitoring, stabilizing, and modulating the solitons. Towards building photonic integrated circuits for next-generation applications, that will simultaneously maximize system performance and minimize size, weight, and power consumption metrics, achieving soliton microcombs and efficient EO modulation on a chip is essential. X-cut thin-film lithium niobate (TFLN) has emerged as a leading photonic platform for the realization of high-performance integrated EO devices and systems. However, despite extensive research, soliton microcombs have remained elusive to X-cut TFLN due to its multiple strong Raman-active modes, in-plane refractive index anisotropy, and photorefractive effects. Here, we address this long-standing challenge and demonstrate versatile soliton microcombs on X-cut TFLN, with repetition-rates spanning from the gigahertz (~26 GHz) up to the millimeter-wave (~0.156 THz) regime. The combs feature exceptional long-term stability, maintaining a direct injection-locked state for over 90 minutes (manually terminated), with repetition-rate phase noise closely tracking that of a high-quality electronic microwave synthesizer. Our finding broadly advances both the fundamental science and practical applications of integrated comb sources by enabling efficient EO modulation and broadband coherent solitons to be monolithically combined on the same chip.**

# Introduction

Integrated photonic systems powered by coherent microcombs are gaining attention because of their potential to accelerate a wide range of scientific and technological advancements[1–10]. In many such systems, mode-locked soliton frequency combs emerging from dissipative nonlinear optical microresonators[11] serve as the fundamental light source. The other metric-defining factor across these system-level demonstrations is the electro-optic (EO) modulation efficiency. For example, soliton-based WDM optical data links and photonic computing cores have information transmission and processing rates that scale directly with the EO modulation speed[8–10]. In precision applications where full stabilization of the soliton microcomb is required, feedback signal generation and actuation often necessitate complex EO modulation apparatus[12,13]. The X-cut thin-film lithium niobate (TFLN) photonic platform promises efficient on-chip EO modulation and has already garnered significant demand in both academic and industrial settings[14–16]. Its ultra-low optical loss, large EO effect, and broad EO bandwidth have enabled integrated EO modulators that far surpass their bulk counterparts[17–19]. Notable examples enabled by TFLN EO technology include time-multiplexed photonic computing[20–22], signal processing[23] and radar sensing[24], as well as unity-efficiency frequency shifting[25] and ultrafast laser tuning[26]. In parallel, the Z-cut TFLN platform has been utilized for high-performance soliton microcombs[27–33], but efficient EO components on this platform is substantially lacking. Advancing chip-scale photonic systems to a new frontier, defined by unprecedented performances, compact form factors, and low power consumption, most straightforwardly requires the realization of soliton microcombs on X-cut TFLN. This would allow for its multiple strong optical nonlinearities to be utilized simultaneously.

Soliton microcombs from X-cut TFLN microresonators, operating in the EO-efficient fundamental transverse-electric-like ($TE_{00}$) mode, face three main challenges. First, in-plane refractive index anisotropy necessitates the coupling between $TE_{00}$ and fundamental transverse-magnetic-like ($TM_{00}$) modes below a cutoff wavelength, causing broadband avoided mode crossings (AMXs). Further, waveguide imperfections introduced by the dry-etch fabrication process of TFLN promotes $TE_{00}$/$TM_{00}$ AMXs even far beyond this cutoff[34]. Such AMXs distort the microresonator's integrated dispersion, known to prevent soliton formation[35]. Second, the photorefractive effect on X-cut TFLN, induced by the optical excitation of charge-carriers, leads to refractive index changes that hinder the ability to maintain a stable pump-cavity detuning[36,37], nevertheless a critical requirement for microcombs. Lastly, X-cut TFLN exhibits multiple strong Raman-active modes that obstruct the initialization pathway for cavity soliton formation[38]. For these reasons, nearly half a decade of soliton research has been limited to Z-cut TFLN[27–33] owing to its in-plane isotropic refractive index and unique photorefractive effect facilitating self-starting soliton microcombs[39]. Moreover, parasitic Raman lasing due to the ~630 cm$^{-1}$ Raman mode on Z-cut TFLN was successfully suppressed through the use of specialized coupling structures[33,40]. This approach, however, is ineffective in the case of X-cut TFLN, where the primary Raman shift experienced by the $TE_{00}$ mode is too small (~251 cm$^{-1}$). Another approach to suppress

Raman lasing based on rotating resonator orientations has been demonstrated on X-cut TFLT (tantalate), but instead the corresponding amount of suppression was found to be insufficient when implemented on X-cut TFLN[41].

In this work, we overcome the long-standing challenge and achieve soliton microcombs on X-cut TFLN featuring gigahertz (~ 26 GHz) and millimeter-wave (up to ~0.156 THz) repetition-rates. Moreover, we demonstrate direct injection-locking of a gigahertz repetition-rate soliton utilizing EO phase modulation, effectively disciplining its repetition-rate phase noise to match that of the driving microwave source[42–44]. This injection-locked state is maintained over a long-term, resulting in long-term stability and three orders of magnitude longer uptime than current demonstrations on X-cut TFLT[41]. The successful soliton generation on X-cut TFLN is supported by three advances. First, using confocal Raman spectroscopy, we show more than an order-of-magnitude variation in the Raman gain depending on the polarization of the excitation laser. This finding motivates racetrack resonators designed to feature a large $TE_{00}$ mode fraction aligned with the extraordinary z-axis, thus minimizing the round-trip averaged Raman gain experienced by each resonance mode. Second, when optically pumping such resonators, we show that they do not exhibit photorefraction-dominated cavity dynamics and are instead governed by effects consistent with thermo-optic self-stability[45]. Third, we regulate the strength of birefringence-induced AMXs by adjusting the waveguide bending geometry, which in turn modulates the waveguide roughness-mediated $TE_{00}$/$TM_{00}$ coupling[34]. To this end, we demonstrate that soliton microcomb generation remains robust against moderate AMXs.

## Results

### *X-cut TFLN soliton generator*

Our devices are fabricated on X-cut TFLN-on-insulator chips (Fig. 1**(a)**, **(b)**), which were cleaved from a commercially available wafer (NANOLN). Prior to fabrication, we performed polarization-resolved confocal Raman spectroscopy on bare chips and found that the extent of Raman scattering displays a strong dependence on the incident excitation's polarization (Fig. 1**(c)**), in contrast to a previous study[41]. When this polarization is aligned with the ordinary y-axis of the crystal ($\theta = 90$), we observed a greater than 10 dB reduction in peak gain for the 250.5 cm$^{-1}$ Raman mode, compared to when the polarization is aligned with the extraordinary z-axis ($\theta = 0$). We note that the quoted Raman shift is not absolute, as the spectrometer configuration is limited to a measurement resolution of 2 cm$^{-1}$ (see Supplementary information for details). For effective suppression of Raman lasing, we designed several racetrack microresonators with straight propagating sections oriented along the z-axis, ensuring that the electromagnetic field vector of $TE_{00}$ light is predominantly aligned with the y-axis. The bending transitions between y- and z-axes utilize complete Euler curves defined by a variable minimum bending radius (BR).

To investigate the challenges of Raman lasing and the photorefractive effect in the context of soliton microcomb generation on X-cut TFLN, we first characterized a ~26 GHz-FSR microresonator (see Supplementary information for device parameters). In our characterization, we first tuned the pump laser frequency to linearly scan back and forth within approximately 3 GHz of typical $TE_{00}$ modes of the resonator at a scan rate of 30 Hz, which is sufficiently slow to capture the quasi-steady intracavity thermal and photorefractive effects. All the resonance modes' transfer functions smoothly transitioned from symmetric Lorentzians in the forward and backward scan to a hysteretic behavior consistent with thermo-optic self-stability[45]. One example is shown in Fig. 2**(a)**, as the optical power on-chip is increased in even steps of 3 mW. None of the scans exhibit rapid switching between on- and off-resonance states[36], or broad triangular shapes in the red-to-blue scanning direction[37], indicating the absence of photorefraction-dominated dynamics in our device. Pumping this resonance with approximately 33 mW of optical power on-chip (far above the parametric oscillation threshold), we step through the blue-detuned regime and confirm the pump-cavity self-stability under blue detuning as well as record the canonical pathway towards soliton mode-locking (Fig. 2**(b)**). We observed the formation of primary combs, secondary-combs, and merging sub-combs, which provide the critical initial conditions for the formation of cavity solitons. Importantly, we found that Raman lasing signatures were completely absent in the measured optical spectra, despite the small FSR of the microresonator. These results provide key evidence that X-cut TFLN resonators, designed and fabricated in our approach, are compatible with a wide range of established methods for the generation, stabilization, and manipulation of soliton microcombs.

### *Soliton state space in a GHz-FSR X-cut TFLN microresonator*

We adopt the counter-propagating bichromatic pumping scheme for soliton generation[46,47]. A simplified experimental setup is illustrated in Fig. 3**(a)**. Here, an L-band $TE_{00}$ resonance is pumped by the "main" pump, and a C-band $TM_{00}$ resonance is pumped by the "cooler" pump. Essentially, the cooler pump provides intracavity thermal balancing as the nonlinear comb state transitions from the high-power modulation instability state to the low-power soliton state. We chose to cool in a different polarization family so as to minimize cross-phase modulation between the cooler pump and the main pump, as well as the soliton. When a thermally-balanced condition is achieved, a large number of soliton steps are revealed. Traversing the soliton state space is stochastic, as illustrated by distinct soliton step sequences in the ten comb power traces collected 1 s apart (Fig. 3**(b)**). Due to the detuning stability and extended soliton access range provided by the interplay between the main and cooler pumps, we were able to achieve adiabatic tuning and on-demand synthesis of soliton pulses in a rich state space, with the six lowest-order soliton states shown in Fig. 3**(c)**, **(d)**. We observed both 6- and 2-perfect soliton crystal states, characterized by large repetition-rates of ~156.42 and ~52.14 GHz, respectively. The perfect crystal nature is evidenced by the vanishing microwave beat note at the fundamental FSR frequency (see Supplementary information). The microresonator supporting such states features an integrated dispersion following a strictly upward-opening parabola, indicating strong anomalous dispersion at the main pump's location. We note that $TE_{00}/TM_{00}$ AMXs in this device are very weakly perturbative. Accordingly, the single-soliton envelope is highly predictable and closely follows the theoretical $sech^2$ shape, with only local deviations mapping to AMX locations.

In the preceding discussion, the main pump's resonance mode features a high intrinsic quality factor ($Q_i$) of $4.8 \cdot 10^6$ (Fig. 3**(e)**). The soliton number for each state is determined by their step-wise distributed total comb power (Fig. 3**(f)**). For the single-soliton state, we utilized a 43 GHz-bandwidth fast photodetector (PD1) and electronic spectrum analyzer (ESA) to directly synthesize and measure the microwave-rate beatnote given by 26.0846 GHz (Fig. 3**(g)**).

*Injection-locking and long-term stability of single-soliton*

The long-term stability of solitons plays a decisive role in determining their viability when deployed in diverse, system-level settings. In applications such as microwave distribution over fiber using solitons, maintaining a low-noise soliton repetition-rate is crucial, though this is often challenging due to pump laser fluctuations that couple into the repetition-rate noise. To evaluate both aspects of the GHz-repetition-rate single-soliton state, we apply EO phase modulation to the pump which effectively traps the soliton to the modulation. Such direct injection-locking has two important consequences, which we demonstrate the spectral purification and tuning of the soliton repetition-rate in Figs. 4**(a)** to **(c)**. When the single-soliton is free-running (continuous wave pump), the repetition-rate drifts. Conversely, when it is injection-locked by the phase modulation (modulation depth ~$0.2\pi$), the repetition-rate is exactly pinned to the modulation frequency and the beatnote exhibits significant linewidth reduction. Indeed, we find that the beatnote's single-sideband (SSB) phase noise, measured from 100 Hz to 1 MHz offset frequencies, reduces down to the source level. The bump near 1 kHz offset, present in both free-running and injection-locked cases, is likely due to residual noise from the pump laser, which is not stabilized. Next, we linearly sweep the modulation frequency over a span of 600 kHz and find an injection-locking range of ~107 kHz, over which the repetition-rate adiabatically follows the modulation frequency sweep, demonstrating dynamic tunability. Remarkably, the injection-locked state shown here is preserved over very long time-scales, exceeding 90 minutes and only manually terminated. We collect soliton spectra throughout the measurement and stack them in Fig. 4**(d)**. We note that up to the spectrum analyzer resolution (0.02 nm, or ~2.5 GHz), comb lines of distinct spectra perfectly overlap throughout the entire soliton bandwidth and for all times, indicating high quality of the injection lock. In other words, pump laser fluctuations do not induce offsetting comb lines through repetition-rate fluctuations, even if propagated to large mode numbers. The repetition-rate beatnote is continuously monitored and remains stable. Two "wings" around the beatnote correspond to the bump near 1 kHz offset in Fig. 4**(b)**.

## Multi- and single-solitons in mmWave-FSR X-cut TFLN microresonators

Further study is required regarding scaling down the resonator size to realize mmWave repetition-rate soliton microcombs. In this regime, taming the birefringence-induced $TE_{00}/TM_{00}$ coupling becomes critical, since excessive AMXs perturbing the microresonator's integrated dispersion can prevent soliton formation[35]. The ~26 GHz-FSR resonator utilizes a minimum BR of 80 $\mu$m, which we found to be large enough to mitigate significant perturbations. In shallow-etched waveguides, the choice of BR can strongly influence the mixing between $TE_{00}$ and $TM_{00}$ modes. This mixing occurs as the refractive index of the $TE_{00}$ mode transitions from the ordinary to the extraordinary axis through the bends, and it has its origin in the in-plane refractive index anisotropy but can be further exacerbated by scattering mechanisms (primarily waveguide sidewall roughness). Utilizing this fact, the extent of $TE_{00}/TM_{00}$ coupling can be controlled by varying the minimum BR. Larger minimum BR leads to more adiabatic transitions, while the optical mode is more symmetric over the waveguide cross section and experiences less overlap with the waveguide's outer sidewall. On the contrary, a smaller minimum BR results in more abrupt transitions, exacerbating the coupling between $TE_{00}/TM_{00}$ and results in periodically occurring, broad ranges of AMX-affected modes. Practically, the minimum BR also cannot be increased indefinitely, as it fundamentally limits achievable resonator FSRs and compromises the ordinary to extraordinary index fraction (Fig. 4**(a)**). Without extra consideration in the fabrication of X-cut TFLN soliton microcomb generators, these aspects set an upper bound for single-soliton repetition-rates.

To investigate this, we fabricated on a separate chip a series of racetrack microresonators with substantially smaller minimum BRs of 60 $\mu$m (see Supplementary information for device parameters). Comparison of these devices with the 80 $\mu$m-BR device enables an examination of the impact of varying birefringence-induced mode-mixing on the existence of soliton states. All resonators feature a relatively high $Q_i^2 \cdot FSR$ (median value, Fig. 4**(b)**), which is inversely proportional to the four-wave-mixing parametric oscillation threshold. As shown in Fig 4**(c)**, the 60 $\mu$m-BR microresonators exhibit overall much stronger AMXs in their measured integrated dispersions (compared to Fig. 3**(d)**). Still, we could generate adiabatically-tuned multi- and single-soliton states with mmWave repetition-rates (Fig. 4**(d)**). While multiple AMXs of moderate strengths are imprinted as broad Fano-like structures onto the soliton envelopes, they did not prevent soliton existence or free-running operation. As an important note, our devices are fabricated using an optimized workflow that incorporates angled etching[48], which significantly reduces the redeposition profile immediately post-etching. This minimizes the directionally-dependent wet etch (RCA-1) processing[49], which critically enables a symmetric waveguide cross section and further reduces $TE_{00}/TM_{00}$ coupling strength (see Supplementary information for fabrication details).

## Discussion

We explored the design space for microresonators that host cavity solitons on the X-cut TFLN platform. We established and experimentally verified insights towards Raman lasing suppression, achieving thermo-optic-dominated cavity nonlinear response, and controlling birefringence-induced perturbations in the microresonator's integrated dispersion. Such insights enabled our demonstration of multi- and single-soliton states featuring GHz- up to mmWave-repetition rates on X-cut TFLN. The fabricated resonators could be improved in a few ways. First, increasing the etching depth can reduce the minimum BR, thus enabling smaller device footprint while still maintaining low loss[30]. Second, reducing the initial film stack from 600 nm can further push the $TE_{00}$/$TM_{00}$ hybridization cutoff to lower wavelengths, thus reducing the microresonators' integrated dispersion perturbations near the telecommunication bands[34] and allow for larger soliton bandwidths and repetition-rates. Third, the microresonator dispersion can be engineered for dispersive wave emission[50] or flat-top comb spectra[51] in the small anomalous to near-zero dispersion regimes. We evaluated the long-term stability and direct injection-locking capability of the GHz-repetition-rate single-soliton, unequivocally confirming its usability in practical system-level experiments, spanning applications simply requiring continuous soliton uptime to those demanding high soliton spectral purity. We note that the single EO phase modulator can be integrated on the X-cut TFLN chip, supporting multiple functions not only including injection-locking, but also Pound-Drever-Hall stabilization to the microresonator[44] as well as repetition-rate measurement and feedback[52,53]. In fact, state-of-the art X-cut TFLN phase modulators can already feature π-phase-shift voltages below 2 V at ~26 GHz and would in principle achieve zero insertion loss between the modulator and the soliton generator[54].

Overall, we conclude that the soliton dynamics supported by Raman-suppressed X-cut TFLN racetrack microresonators are consistent with the existing knowledge on soliton microcombs[11]. This has several implications. Operationally, conventional soliton triggering methods are expected to be effective, including fast frequency tuning via suppressed-carrier single-sideband EO modulators[55] and rapid thermo-optic heating[56]. In fact, the necessary modulators[18] and heaters[57] could be integrated directly on the X-cut TFLN chip as well, reducing the system complexity even further. At the device level, we anticipate that our designs would enrich the toolbox of comb sources available to X-cut TFLN, including Kerr-effect-based optical parametric oscillators[58], meta-dispersion[59] and photonic molecule[60] soliton microcombs, as well as dark-pulses[61,62]. Considering the applications, novel synchronization phenomena for chip-based optical clocks[12,63], analog simulators of time-dependent quantum systems[64], optimizers based on coherent Ising machines[65], as well as massively parallel WDM communications[8–10], ranging[66] and signal processing[5,6], could all benefit from soliton microcombs in conjunction with large-scale EO modulator networks and efficient second-order nonlinear optics[67] on X-cut TFLN. As this platform has already achieved commercialization and supports multi-project wafer designs through specialized TFLN foundries, our result places monolithically integrated, soliton-driven X-cut TFLN photonic systems within practical reach.

**Acknowledgements:** Y.S. thanks Nicholas Achuthan, Rui Jiang, Leticia Magalhães, Dylan Renaud, Urban Senica, Pradyoth Shandilya, Linbo Shao, Neil Sinclair, Chenjie Xin, Zhiquan Yuan, Tianyi Zeng, and Chaoshen Zhang for discussions related to all aspects of this project. Y.S. thanks Jiayu Yang and Anna Shelton for respectively providing an MgO: X-cut TFLN sample and an X-cut TFLT sample. Y.S. thanks Arthur McClelland for instruction on Raman spectroscopy and Jiangdong Deng for equipment advice. Y.S. and M.L. thank Ganna Savostyanova and Meg Reardon for administrative support. G.H. acknowledges support from the Swiss National Science Foundation Postdoc Mobility Fellowship. The device fabrication in this work was performed at the Harvard University Center for Nanoscale Systems (CNS); a member of the National Nanotechnology Coordinated Infrastructure Network (NNCI), which is supported by the National Science Foundation under NSF award no. ECCS-2025158.

**Author contributions**: M.L. and Y.S. conceived the project. Y.S. developed the fabrication process, designed the devices, performed the experiments, and analyzed the data. Y.S. and X.Zhu fabricated the devices. X.Zuo provided advice on the fabrication. G.H. provided advice on the experiment. Y.S. and M.L. wrote the manuscript with comments from all authors. M.L. supervised the project.

**Funding**: Air Force Office of Scientific Research (FA955024PB004), National Science Foundation (ECCS-2407727), Naval Air Warfare Center Aircraft Division (N6833522C0413), National Research Foundation funded by the Korea government (NRF-2022M3K4A1094782).

**Competing interests**: M.L. is involved in developing lithium niobate technologies at HyperLight Corporation. The authors declare no other competing interests.

**Data and materials availability**: All data needed to evaluate the conclusions in the paper are present in the paper and/or the Supplementary Information.

**Additional note:** During the manuscript preparation stage, we became aware of another preprint reporting soliton microcombs on X-cut TFLN[68].

67. McKenna, T. P. *et al.* Ultra-low-power second-order nonlinear optics on a chip. *Nat. Commun.* **13**, 4532 (2022).

68. Nie, B. *et al.* Soliton microcombs in X-cut LiNbO$_3$ microresonators. Preprint at https://doi.org/10.48550/arXiv.2502.07180 (2025).

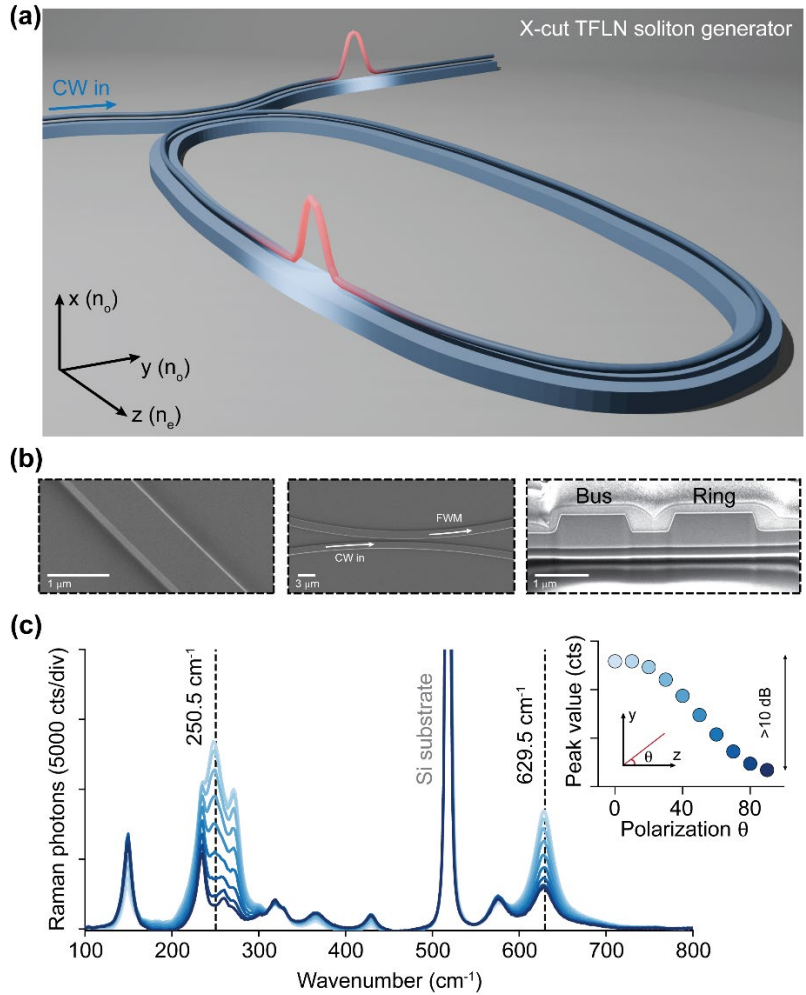

**Fig. 1 | Raman-suppressed X-cut TFLN racetrack microresonators for soliton microcomb generation. (a)** Schematic of the X-cut TFLN soliton generator, based on a racetrack microresonator whose straight waveguide section is propagating along the $z$-axis. **(b)** Scanning electron microscopy images of the optical circuit. Left panel: a high-quality tapered-waveguide bend with smooth sidewalls. Middle panel: top-down view of the nearly-single-mode-supporting coupling region. Right panel: focused ion beam produced-image of a 0.7 $\mu$m coupling gap. Note that the waveguides are shallow-etched using an optimized angled-etching process that yields symmetric sidewalls with ~74 degrees sidewall angle. See Supplementary information for elaborations on the microresonator design. **(c)** Polarization-dependent Raman spectrum of X-cut TFLN collected prior to fabrication of the chip. Raman-active modes are marked by the black dashed lines, and the peak near 520 cm$^{-1}$ is due to the silicon substrate. The inset shows more than 10 dB suppression in the peak Raman intensity for the 250.5 cm$^{-1}$ mode, when the excitation polarization is aligned with the ordinary $y$-axis compared to the extraordinary $z$-axis. Vertical scale of the inset is identical to that of **(c)**.

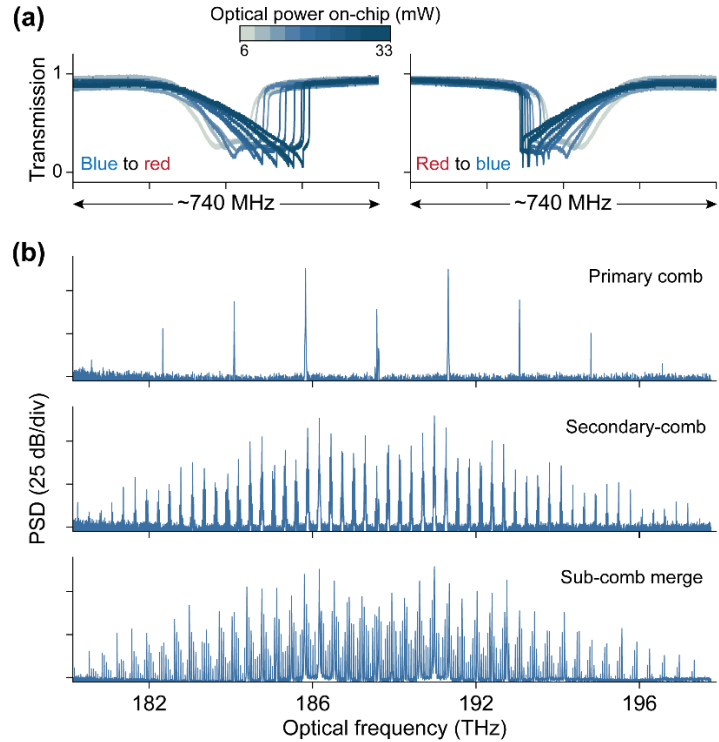

**Fig. 2 | Microresonator dynamics and blue-detuned nonlinear states. (a)** Pump-cavity detuning scans of an L-band pump across a typical $TE_{00}$ mode. Left panel: blue to red detuning scan, described by a broad triangular shape. Right panel: red to blue detuning scan, described by a sharp spike followed by resonance relaxing back. Ten laser scans are overlaid to illustrate the smooth transition from symmetric Lorentzian shapes (with a small splitting) at low optical pump powers, to a hysteretic behavior governed by thermo-optics at high optical pump powers. Notably, photorefraction-dominated cavity dynamics are completely absent. From light gray to dark blue traces, the optical powers on-chip are calibrated to span from 6 to 33 mW, in even steps of 3 mW. **(b)** Stable, blue-detuned nonlinear states as the detuning is decreased, including primary comb, secondary-comb, and sub-comb merging. There are no spurious oscillations corresponding to Raman lasing. The optical power on-chip corresponds to the dark blue trace in Fig. 2**(a)**, given by 33 mW.

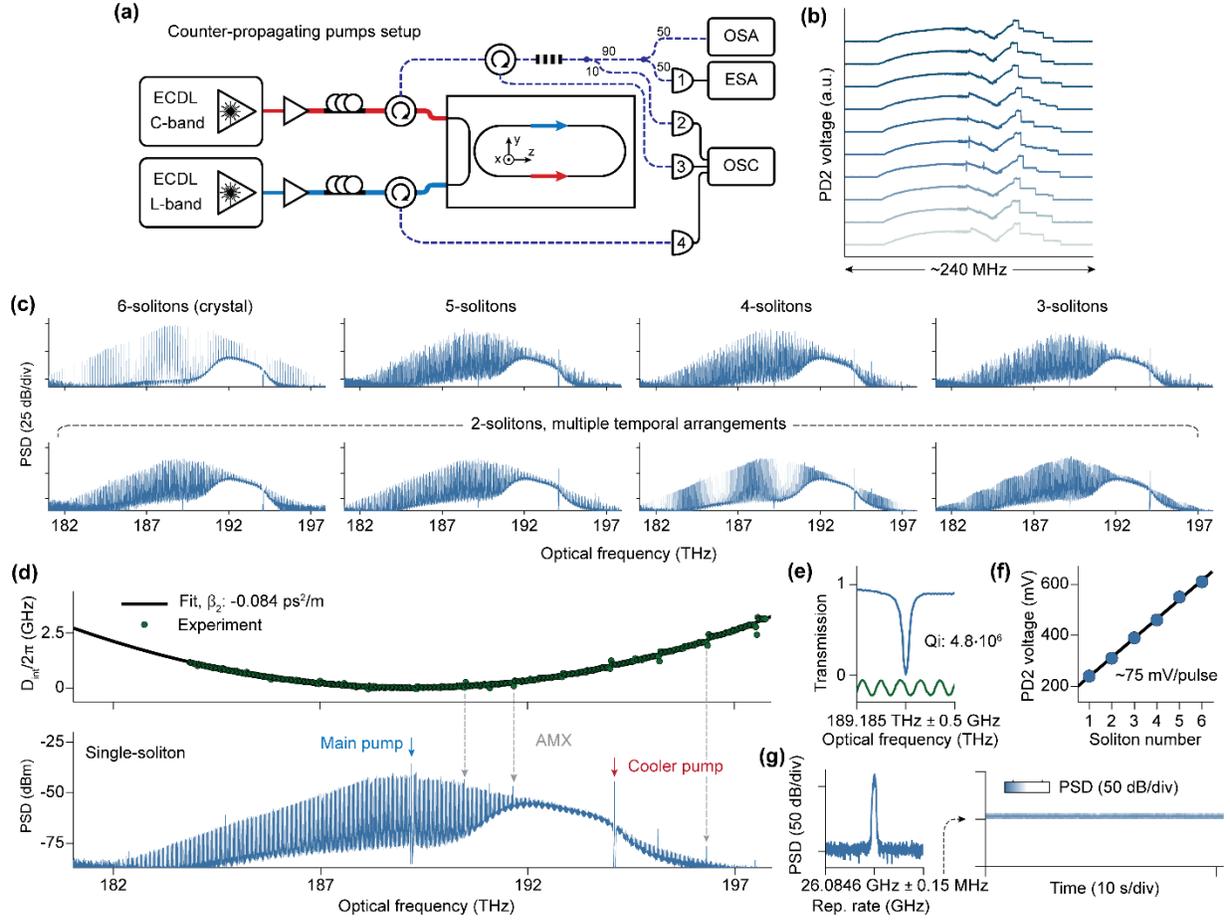

**Fig. 3 | Diverse soliton states from a high-quality, GHz-FSR, X-cut TFLN microresonator.** (a) Counter-propagating bichromatic pumps setup. The main pump at 1584.65 nm (L-band, $TE_{00}$) is forward propagating, and the cooler pump at 1544.72 nm (C-band, $TM_{00}$) is counter-propagating. The amplified spontaneous emission of the main pump's erbium-doped fiber amplifier is removed by a bandpass filter, and any residual, facet-reflected cooler pump in the forward propagating direction is removed by a fiber Bragg grating (both not shown in the schematic here, for simplicity). Numbers label photodetectors (PDs), where 1 is a 43-GHz bandwidth PD (synthesizes the soliton beatnote) and 2-4 are 125-MHz bandwidth PDs (measures the comb power, main pump power, and cooler pump power, respectively). (b) Ten comb power traces collected 1 s apart, illustrating the stochastic nature of soliton steps. (c) Family of multi-soliton states, including perfect 6- and 2-soliton crystal states (first panel, top row and third panel, bottom row) which have repetition-rates of ~0.156 THz and ~52 GHz, respectively and vanishing beatnote at the fundamental microresonator FSR. (d) Microresonator's integrated dispersion and single soliton state. The main (cooler) pump is marked in blue (red), and small AMXs in the integrated dispersion are mapped to local spurs in the soliton's $sech^2$ spectral envelope. (e) The $TE_{00}$ pump mode features a high $Q_i$ of $4.8 \cdot 10^6$, calibrated by a 191.5 MHz-period fiber Mach-Zehnder interferometer (offset green curve). (f) Attenuated comb power detected on PD2 as a DC voltage, which distinguishes the soliton number for each state in this Figure. (g) Single-soliton beatnote directly synthesized by PD1 and measured on an electronic spectrum analyzer (ESA). The measurement range is 0.3 MHz, and the resolution bandwidth (RBW) is 100 Hz.

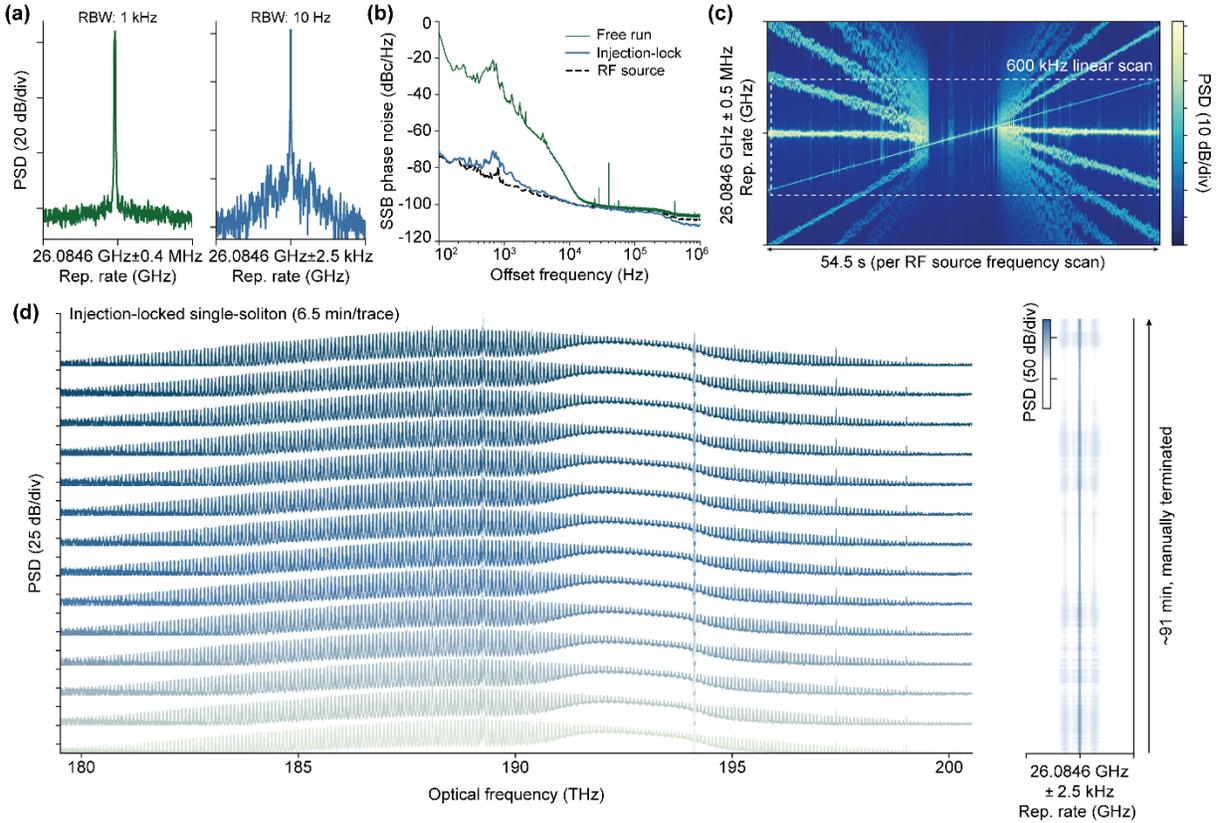

**Fig. 4 | Injection-locking and long-term stability of single-soliton state. (a)**. Single-soliton repetition-rate beatnote synthesized by PD1 and measured on the ESA, in the free-running state (green, left) and in the direct injection-locked state (blue, right). Note that the measurement spans are 0.8 MHz and 5.0 kHz, and the RBWs are 1 kHz and 10 Hz, respectively. **(b)**. SSB phase noise of the repetition-rate beatnote in the free-running state (green) and in the direct injection-locked state (blue), compared to that of the microwave source (black). More than 60 dB reduction in the SSB phase noise is observed at 100 Hz offset frequency, and the beatnote closely mimics the source. **(c)**. Dynamic ESA spectrum as the microwave source is linearly scanned over 600 kHz (dashed white box). Multiple frequencies collapse into a single beatnote over the injection-locking range. The locking range is estimated to be ~107 kHz. **(d)** Left panel: stacked soliton spectra collected sequentially in time. Each spectrum takes about 6.5 minutes for a full bandwidth scan. Right panel: repetition-rate beatnote continuously monitored for over 90 minutes. The measurement span is 5.0 kHz and the RBW is 10 Hz, and the injection-locked state is maintained throughout this time. We note that this state was not lost at the end, rather it was manually terminated.

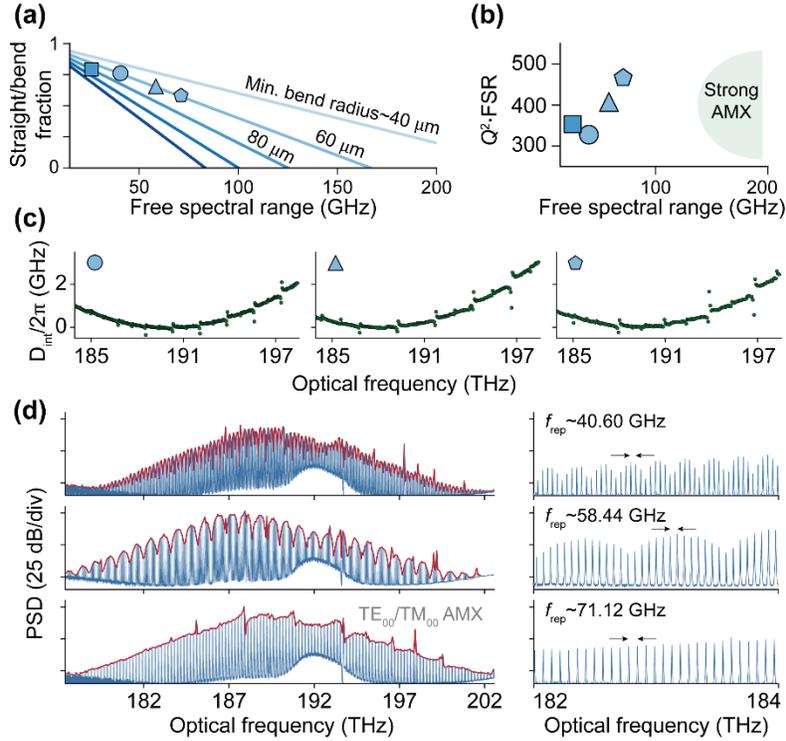

**Fig. 5 | Scaling-down microresonator size for mmWave-repetition-rate solitons. (a)** Inter-dependence between racetrack resonator straight/bend fraction (related to $n_o/n_e$ composition) and its FSR. Lines represent varying minimum BR, spanning from 40 $\mu$m (light blue) to 120 $\mu$m (dark blue), in even steps of 20 $\mu$m. Square marks the ~26 GHz-FSR device with 80 $\mu$m-BR in Fig. 2, 3. Circle, triangle, and pentagon mark the ~41, 58, and 71 GHz-FSR devices with 60 $\mu$m-BR in this Figure. **(b)** Measured median $Q_i^2 \cdot FSR$ as a function of microresonator FSR (size). The green shaded region represents a strong AMX zone for small microresonators with native FSRs in the hundreds of GHz range. This region is accessible only with minimum BR below 40 $\mu$m; however, soliton crystal states from large resonators can also reach this regime of repetition-rates. **(c)** Microresonators' integrated dispersions. The AMX perturbations are enhanced due to the reduced BR and increased surface scattering. **(d)** Multi- and single-soliton states featuring mmWave repetition-rates of 40.60, 58.44, and 71.12 GHz. The left panels display comb spectra in blue, with red-highlighted envelopes serving as a visual guide. Broad Fano-like structures are due to extended AMXs in the integrated dispersion, while the sharp spikes are due to some cross phase modulation between the main and cooler pumps. Right panels provide zoomed-in views of a 2 THz window covering the spectral range of 182 to 184 THz, showcasing the varied repetition-rates and the sub-structure in the spectral envelopes.